\documentclass[tighten,twocolappendix,twocolumn]{aastex631}

\usepackage{graphicx}	% Including figure files
\usepackage{amsmath}	% Advanced maths commands
\usepackage{hyperref}
\newcommand{\uat}[2]{\href{http://astrothesaurus.org/uat/#2}{#1 (#2)}}

\begin{document}

\title{Measuring the Hubble constant through the galaxy pairwise peculiar velocity}

\correspondingauthor{Wangzheng Zhang}
\email{1155129240@link.cuhk.edu.hk}

\author[0000-0003-0102-1543]{Wangzheng Zhang}
\affiliation{Department of Physics, the Chinese University of Hong Kong, Sha Tin, NT, Hong Kong}

\author[0000-0002-1971-0403]{Ming-chung Chu}
\affiliation{Department of Physics, the Chinese University of Hong Kong, Sha Tin, NT, Hong Kong}

\author[0000-0001-7075-6098]{Shihong Liao}
\affiliation{Key Laboratory for Computational Astrophysics, National Astronomical Observatories, Chinese Academy of Sciences, Beijing 100101, China}

\author{Shek Yeung}
\affiliation{Department of Physics, the Chinese University of Hong Kong, Sha Tin, NT, Hong Kong}

\author[0000-0002-1908-0384]{Hui-Jie Hu}
\affiliation{University of Chinese Academy of Sciences, No. 19 A Yuquan Road, Beijing 100049, China}
\affiliation{Key Laboratory for Computational Astrophysics, National Astronomical Observatories, Chinese Academy of Sciences, Beijing 100101, China}

\begin{abstract}
The Hubble constant $H_0$, the current expansion rate of the universe, is one of the most important parameters in cosmology. The cosmic expansion regulates the mutually approaching motion of a pair of celestial objects due to their gravity. Therefore, the mean pairwise peculiar velocity of celestial objects, which quantifies their relative motion, is sensitive to both $H_0$ and the dimensionless total matter density $\Omega_m$. Based on this, using the Cosmicflows-4 data, we measured $H_0$ for the first time via the galaxy pairwise velocity in the nonlinear and quasi-linear range. Our results yield $H_0=75.5\pm1.4\ \mathrm{km}\ \mathrm{s}^{-1}\ \mathrm{Mpc}^{-1}$ and $\Omega_m=0.311^{+0.029}_{-0.028}$. The uncertainties of $H_0$ and $\Omega_m$ can be improved to around 0.6\% and 2\%, respectively, if the statistical errors become negligible in the future.
\end{abstract}


\section{Introduction} \label{sec:intro}
Although the Lambda Cold Dark Matter ($\Lambda$CDM) model achieves great success in describing a wide range of cosmological observations, the Hubble tension, i.e., the significant difference in the values of the Hubble constant $H_0$ between indirect and direct measurements, suggests that physics beyond $\Lambda$CDM may be needed. Specifically, the $H_0$ determined by the anisotropies of cosmic microwave background (CMB) from Planck ($H_0=67.36\pm0.54\ \mathrm{km}\ \mathrm{s}^{-1}\ \mathrm{Mpc}^{-1}$, \citealt{aghanim2020planck}) deviates by almost 5$\sigma$ from that measured using Type Ia supernovae (SNeIa) calibrated by Cepheids (SH0ES, $H_0=73.04\pm1.04\ \mathrm{km}\ \mathrm{s}^{-1}\ \mathrm{Mpc}^{-1}$, \citealt{SHOES2022}). More details can be found, for example, in \cite{review2022_s1,review2022_s2,review2022_s3}. Other independent CMB measurements, such as ACT \citep{ACT2020} and SPT-3G \citep{SPT3G2023}, give consistent results on $H_0$ as Planck's. $H_0$ can also be measured using different distance ladder calibrators other than Cepheids, such as the tip of the red-giant branch \citep[TRGB,][]{TRGB2021_s1, TRGB2022_s2}, surface brightness fluctuations (SBF) of galaxies \citep{SBF2021_s1, SBF2021_s2}, Miras variables \citep{Miras2020}, or the Tully-Fisher relation \citep[TFR,][]{BTFR2020_s1, BTFR2020_s2}. Furthermore, $H_0$ can be obtained from other independent measurements that do not rely on the distance ladder, such as through Masers \citep{Maser2020}, gravitational lensing \citep{H0LiCOW2020,TDCOSMO2020}, gravitational waves \citep{GW2021_s1,GW2023_s2}, or cosmic chronometers \citep{CC2022}. Although not all of the direct measurements of $H_0$ are in tension with indirect measurements, most are significantly larger than the latter, particularly Planck's.

Therefore, more independent ways to measure $H_0$ are desirable. In this Letter, we measure $H_0$ using galaxy velocity statistics in the nonlinear and quasi-linear regime for the first time. Typically, $H_0$ is derived by balancing the overall infall and outflow of galaxies \citep{tully2016,tully2023}. Here, we extend this by considering the relative motion of galaxies, i.e., through the mean pairwise peculiar velocity $v_{12}(r)$, also known as pairwise velocity, which is defined as
\begin{equation}\label{eq:pwv-definition}
    v_\mathrm{12}(r)\equiv\langle [\boldsymbol{v}_1(\boldsymbol{r}_1)-\boldsymbol{v}_2(\boldsymbol{r}_2)]\cdot \boldsymbol{\hat r} \rangle,    
\end{equation}
where $\boldsymbol{r}\equiv\boldsymbol{r}_1-\boldsymbol{r}_2$ is the distance vector between a pair of objects such as galaxies or halos in the comoving frame with peculiar velocities $\boldsymbol{v}_{1/2}$. The average $\langle...\rangle$ is taken over all pairs with separation $r\equiv|\boldsymbol{r}|$, and $v_\mathrm{12}(r)$ depends only on $r$ according to isotropy. The pairwise velocity quantifies the relative motion of pairs of objects with separation $r$. For objects with a conserved number, we have
\begin{equation}
    \frac{v_{12}(r,a)}{Hra}=-\frac{a}{3[1+\Xi(r,a)]}\frac{\partial \bar\Xi(r,a)}{\partial a},
\end{equation}
where $\bar\Xi(r,a)\equiv (3/4\pi r^3)\int_0^r d^3\boldsymbol{y} \Xi(y,a)$ and $\Xi$ is the two-point correlation function. $H=\dot a/a$ is the Hubble parameter and $a$ is the scale factor (\citealp[\S6]{mo2010galaxy}; \citealp[\S71]{peebles2020large}). Therefore, the pairwise velocity contains information on how the two-point correlation evolves. At $r\lesssim$ 0.2 Mpc, due to the stable clustering, $v_{12}(r,a)=-Hra$ \citep[\S6]{mo2010galaxy}. At $r\gtrsim$ 30 Mpc, under the linear approximation, $v_{12}(r,a)=-2Hraf\bar{\Xi}(r,a)/3[1+\Xi(r,a)]$ \citep{juszkiewicz1999dynamics}, where the linear growth rate $f\equiv\mathrm{d}\ln D/\mathrm{d}\ln a\approx\Omega_m^{0.55}$ \citep{linder2005Omega0p55}. $D$ and $\Omega_m$ are the linear growth solution and dimensionless total matter density, respectively. Consequently, $v_\mathrm{12}(r)$ is sensitive to $H_0$ and $\Omega_m$. In this Letter, we focus on the nonlinear and quasi-linear range $r\leq16$ Mpc, which requires N-body simulations for accurate predictions and cannot be adequately described by the linear perturbation theory \citep{Kosowsky2009,bhattacharya2011galaxy,mueller2015constraints1,mueller2015constraints2,jaber2024PRD}. There are at least two advantages using $v_{12}(r)$ to measure $H_0$ from the observational point of view. Firstly, any object with a known peculiar velocity can be used in the measurement. Secondly, any common systematic errors of the measured peculiar velocities, e.g., the peculiar motion of the Milky Way, would be largely cancelled in the pairwise velocity.

The pairwise velocity and its related quantities have been well studied to constrain cosmological parameters such as $\sigma_8$ and $\Omega_m$ \citep{juszkiewicz2000evidence, feldman2003estimate, ma2015constraining}, the local growth rate $f\sigma_8$ \citep{Howlett2017MNRAS, dupuy2019estimation}, dark energy or modified gravity models \citep{bhattacharya2008dark, bhattacharya2011galaxy, mueller2015constraints1, bibiano2017pairwise, jaber2024PRD}, and the kinematic Sunyaev-Zeldovich (kSZ) effect \citep{bhattacharya2007cosmological, calafut2021atacama}, which is also used to constrain neutrino masses \citep{mueller2015constraints2}.

The remainder of this Letter is organized as follows. In Sections \ref{sc:simu_method} and \ref{sc:ggpwv}, we discuss how $v_{12}(r)$ is calculated in N-body simulations and observations, respectively. The fitting method and results are elaborated in Section \ref{sc:fitting}. In Section \ref{sc:summary}, we present the summary and discussions.

\section{Halo-halo pairwise velocity}\label{sc:simu_method}

We perform our cosmological N-body simulations using a modified grid-based neutrino-involved \texttt{Gadget-2} \citep{springel2005cosmological,zhang2024MNRAS} with three types of neutrinos with a degenerate mass of 0.02 eV. We sample the Hubble constant $H_0=$ 67.6, 70.6, 73.6, 76.6, and 79.6 $\mathrm{km}\ \mathrm{s}^{-1}\ \mathrm{Mpc}^{-1}$ and $\Omega_m=$ 0.274, 0.304, and 0.334. The scalar index $n_s=0.9652$ and the corresponding spectral amplitude $A_s=2.0968\times10^{-9}$ are from the Planck CMB results \citep{aghanim2020planck}. The baryon density $\Omega_bh^2=0.02244$ is from the big bang nucleosynthesis results \citep{bbnPDG2024}. The initial conditions of the simulations are generated by \texttt{2LPTic} \citep{crocce2006transients}, and the initial power spectra at redshift $z=99$ are obtained by \texttt{CAMB} \citep{Lewis:1999bs}. 

We run dark-matter-only (DMO) simulations with the number of dark matter particles $N_p=1024^3$ and box size $L_{\mathrm{box}}=250\ h^{-1}\mathrm{Mpc}$ ($h\equiv H_0/100$ is the dimensionless Hubble constant). For the set $H_0=73.6\ \mathrm{km}\ \mathrm{s}^{-1}\ \mathrm{Mpc}^{-1}$ and $\Omega_m=0.334$, we set up three runs, which differ only by the initial random seeds. Each run comprises four separate simulations (normal, paired, fixed, paired and fixed), i.e., a total of 12 simulations, to reduce the impact of cosmic variance \citep{paired_and_fixed_2016MNRAS, paired_and_fixed_2020MNRAS}. 

The dark matter halos are found by \texttt{ROCKSTAR} \citep{behroozi2012rockstar}. This Letter adopts halo masses as per \cite{bryan1998ApJ}. The center and velocity of the halo are determined by averaging the positions of dark matter particles in the inner subgroup and the velocities of dark matter particles within the innermost 10\% of the halo's virial radius, respectively \citep{behroozi2012rockstar}. Hereafter, we only consider host halos with at least 200 dark matter particles. The mean halo-halo pairwise peculiar velocity $v_\mathrm{hh}(r)$ is calculated via Eq. (\ref{eq:pwv-definition}). To consistently compare with observations, we convert the default unit $h^{-1}$Mpc in each simulation to Mpc using the corresponding $h$.

\section{Galaxy-galaxy pairwise velocity}\label{sc:ggpwv}

In this Letter, we use the grouped catalog from Cosmicflows-4 (CF-4, \citealt{tully2023}), which minimizes distance uncertainties through weighted averaging of galaxy members. Although our analysis is based on the grouped catalog, we have confirmed that using the brightest galaxies in the ungrouped catalog produces consistent results. The farthest galaxy used in this Letter is at redshift around $0.1$. Hence, we compare the data with only the redshift-0 halos from simulations. As shown in \cite{zhang2024MNRAS}, the magnitude and shape of $v_{\rm{hh}}(r)$ are sensitive to the range of halo virial masses used in the calculation. To consistently compare the data with our simulation results, we need to match their halo mass ranges. Therefore, next, we will discuss how we extract the halo mass information from the CF-4 grouped catalog, followed by the galaxy-galaxy pairwise velocity $v_{\rm{gg}}(r)$ calculation along with its error estimation.

\subsection{Mass information}\label{sc:catalog}

We obtain the stellar masses of the CF-4 galaxies through the light-to-stellar-mass relation \citep{bell2003},
\begin{equation}\label{eq:light-to-stellar-mass}
    \log_{10}\left(\frac{M_g/M_\odot}{L_\lambda/L_{\odot\lambda}}\right)=a_\lambda+b_\lambda\times(B-V),
\end{equation}
where $\lambda$ denotes the different observational bands, and $M_g$ is the galaxy stellar mass. $L_\lambda$ is the luminosity for the $\lambda$ band, and $B-V$ is the difference between the apparent magnitudes in the $B$ and $V$ bands of the galaxy. The coefficients $a_\lambda$ and $b_\lambda$ are taken from Table 7 of \cite{bell2003}. Specifically, we use $B$ and $I$ bands, i.e., $a_B=-0.942$, $b_B=1.737$ and $a_I=-0.399$, $b_I=0.824$ \footnote{In calculation, we subtract $a_\lambda$ by 0.15 dex for the Kroupa initial mass function (IMF) as suggested in \cite{bell2003}.}. The corresponding $B$ and $I$ band luminosities are obtained by
\begin{equation}
    2.5\log_{10}\left(\frac{L_{B(I)}}{L_{\odot B(I)}}\right)
    =M_{\odot B(I)}-m_{B(I)}+\mu_g,
\end{equation}
where the $B$ and $I$ band apparent magnitudes $m_{B(I)}$ and distance moduli $\mu_g$ are taken from HyperLeda \citep{hyperleda2014} after matching with the Principal Galaxies Catalog (PGC) identification. The absolute magnitude of the Sun in the $B$($I$) band is $M_{\odot B (I)}= 5.44\ (4.10)$ \citep{willmer2018}.

Then, we can infer the virial masses of the corresponding host halos $M_h$ from the stellar-to-halo-mass relation \citep{moster2010},
\begin{equation}\label{eq:stellar-to-halo-mass}
	\frac{M_g}{M_h}=2C_0\left[\left(\frac{M_h}{M_{h1}}\right)^{-\beta}+\left(\frac{M_h}{M_{h1}}\right)^{\gamma}\right]^{-1},
\end{equation}
where $C_0=0.02817$, $\beta=1.068$, $\gamma=0.611$ and $M_{h1}=10^{11.899}M_\odot$. 

For each galaxy in the CF-4 grouped catalog, we identify its members using the CF-4 ungrouped catalog and compute the total stellar mass via Eq. (\ref{eq:light-to-stellar-mass}), which is then used to estimate the host halo virial mass via Eq. (\ref{eq:stellar-to-halo-mass}). We find that most of the corresponding host halo virial masses are within the range $[10^{11},10^{13}]M_\odot$. Therefore, we select galaxies and simulation host halos that have host halo virial masses within the range $[10^{11}, 10^{13}] M_\odot$, obtaining a total of around 30,000 galaxies and 200,000 simulation host halos.

\subsection{Peculiar velocity}\label{sc:vp}

Following \cite{davis2014,ma2015constraining}, the line-of-sight peculiar velocity $v_p$ of a certain galaxy is calculated by
\begin{equation}\label{eq:peculiar-velocity}
    v_p=\frac{f(\bar z)V_\mathrm{cmb} - H_0 D_L}{f(\bar z) + H_0 D_L/c},
\end{equation}
where $c$ is the speed of light. The systemic velocity of the galaxy in the CMB rest frame $V_\mathrm{cmb}$ and its luminosity distance $D_L$ are from the CF-4 grouped catalog \citep{tully2023}, while $f(\bar z)$ can be calculated assuming flat $\Lambda$CDM by 
\begin{equation}\label{eq:vcmod}
    f(\bar z)=1+\frac{1}{2}[1-q_0]\bar z-\frac{1}{6}[2-q_0-3q_0^2]{\bar z}^2,
\end{equation}
where $q_0=(3\Omega_m-2)/2$. Given $\Omega_m$ and $H_0$, the cosmological redshift $\bar z$ is calculated from
\begin{equation}\label{eq:zbar}
    D_L(\bar z)=(1+\bar z)\int_0^{\bar z} \frac{c dz'}{H(z')},
\end{equation}
where $H(z)=H_0\sqrt{\Omega_m(1+z)^3+1-\Omega_m}$.

Due to the uncertainties of the distances, the peculiar velocities calculated from Eq. (\ref{eq:peculiar-velocity}) can reach 10,000 km s$^{-1}$ for some galaxies. To exclude these extreme cases, we calculate the line-of-sight velocities of host halos with comparable masses ($[10^{11},10^{13}]\ M_\odot$) in the \texttt{Uchuu} simulations \citep{ishiyama2021Uchuu}. By placing observers at random locations within the simulation box, we generate three mock samples of host halo line-of-sight velocities, all of which follow perfect Gaussian distributions with $3\sigma \approx 1000$ km s$^{-1}$. Therefore, we only consider galaxies satisfying $|v_p|\leq v_p^\mathrm{max}$, with $v_p^\mathrm{max}=1000$ km s$^{-1}$, leaving us with 10,014 galaxies. More details can be found in Appendix \ref{sc:vp_cut}.

\subsection{Position}

The comoving position of each galaxy is calculated by 
\begin{equation}\label{eq:gr_pos}
    \begin{aligned}
        x &= D_c \cos{\delta}\cos{\alpha} ,\\
        y &= D_c \cos{\delta}\sin{\alpha} ,\\
        z &= D_c \sin{\delta},
    \end{aligned}
\end{equation}
where the right ascension J2000 $\alpha$ and declination J2000 $\delta$ of each galaxy are from the CF-4 grouped catalog \citep{tully2023}, and the comoving distance $D_c \equiv D_L/(1+\bar z)$.

\subsection{Galaxy-galaxy pairwise velocity}\label{sc:sub_ggpwv}

As we can only observe the line-of-sight peculiar velocity, we calculate the galaxy-galaxy pairwise velocity by the following estimator \citep{ferreira1999},
\begin{equation}\label{eq:galaxy_pwv}
    v_\mathrm{gg}(r)=\frac{\sum_{A,B} (v_{p,A}-v_{p,B})p_{AB}}{\sum_{A,B} p_{AB}^2},
\end{equation}
where $v_{p,A/B}$ is the line-of-sight peculiar velocity of galaxy $A/B$ calculated using Eq. (\ref{eq:peculiar-velocity}), and $p_{AB}\equiv \boldsymbol{\hat r}\cdot (\boldsymbol{\hat r}_A + \boldsymbol{\hat r}_B)/2$, with $\boldsymbol{r}=\boldsymbol{r}_A-\boldsymbol{r}_B$. The comoving positions of galaxies are denoted as $\boldsymbol{r}_{A/B}$ and calculated using Eq. (\ref{eq:gr_pos}). The summation is over all galaxy pairs with separation equal to $r$. 

Based on Eq. (\ref{eq:galaxy_pwv}), we explore how the variation of $v_p^\mathrm{max}$ (from 1000 km s$^{-1}$ to 800 and 1200 km s$^{-1}$) impacts $v_\mathrm{gg}$ and find that only for $r\lesssim16$ Mpc, $v_\mathrm{gg}$ is not sensitive to the choice of $v_p^\mathrm{max}$. Consequently, we calculate $v_\mathrm{hh}$ and $v_\mathrm{gg}$ in the range $r\in[0,16]$ Mpc using 8 linear bins. More details are provided in Appendix \ref{sc:vp_cut}.

\subsection{Observational error sources}\label{sc:error_analysis}

\begin{figure*}
  \centering
  \includegraphics[width=.45\textwidth]{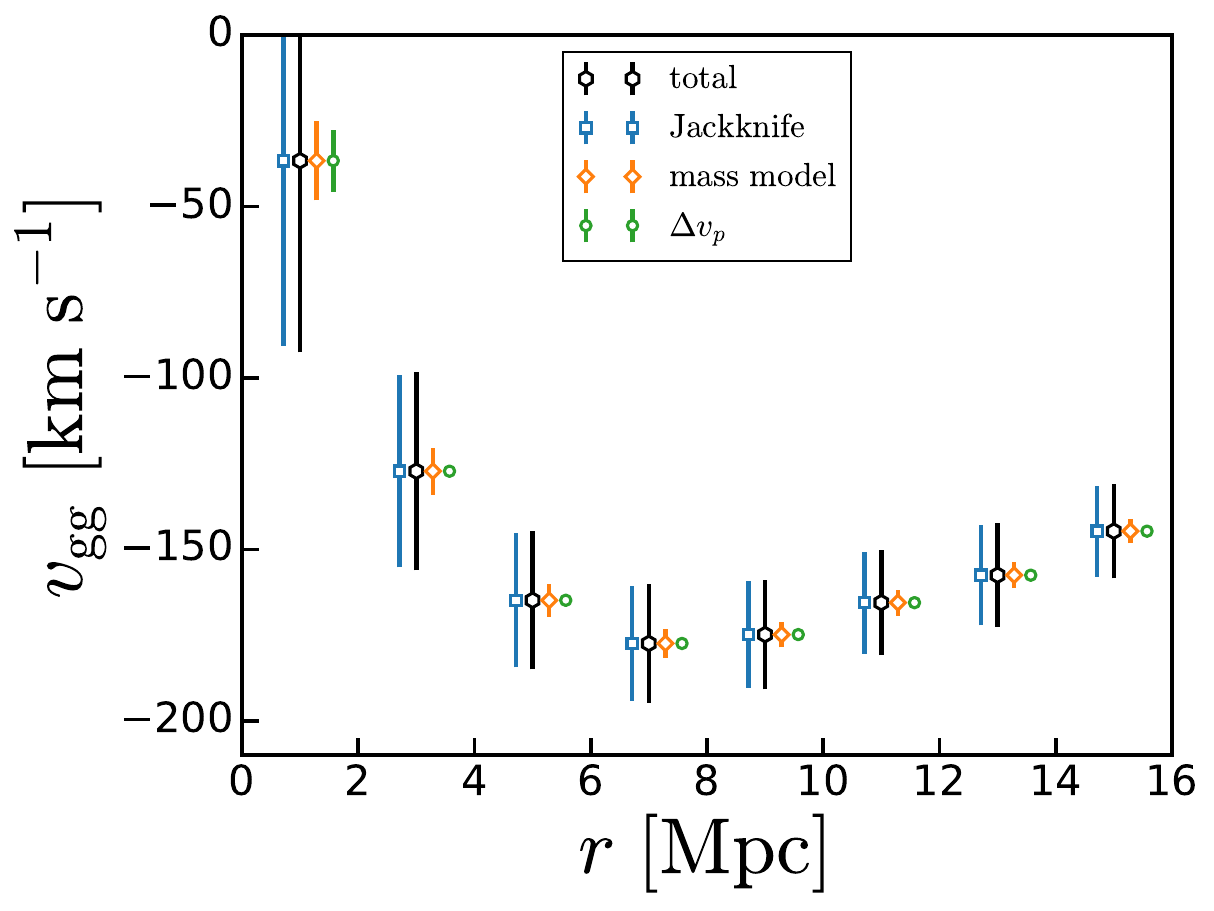}
  \includegraphics[width=.41\textwidth]{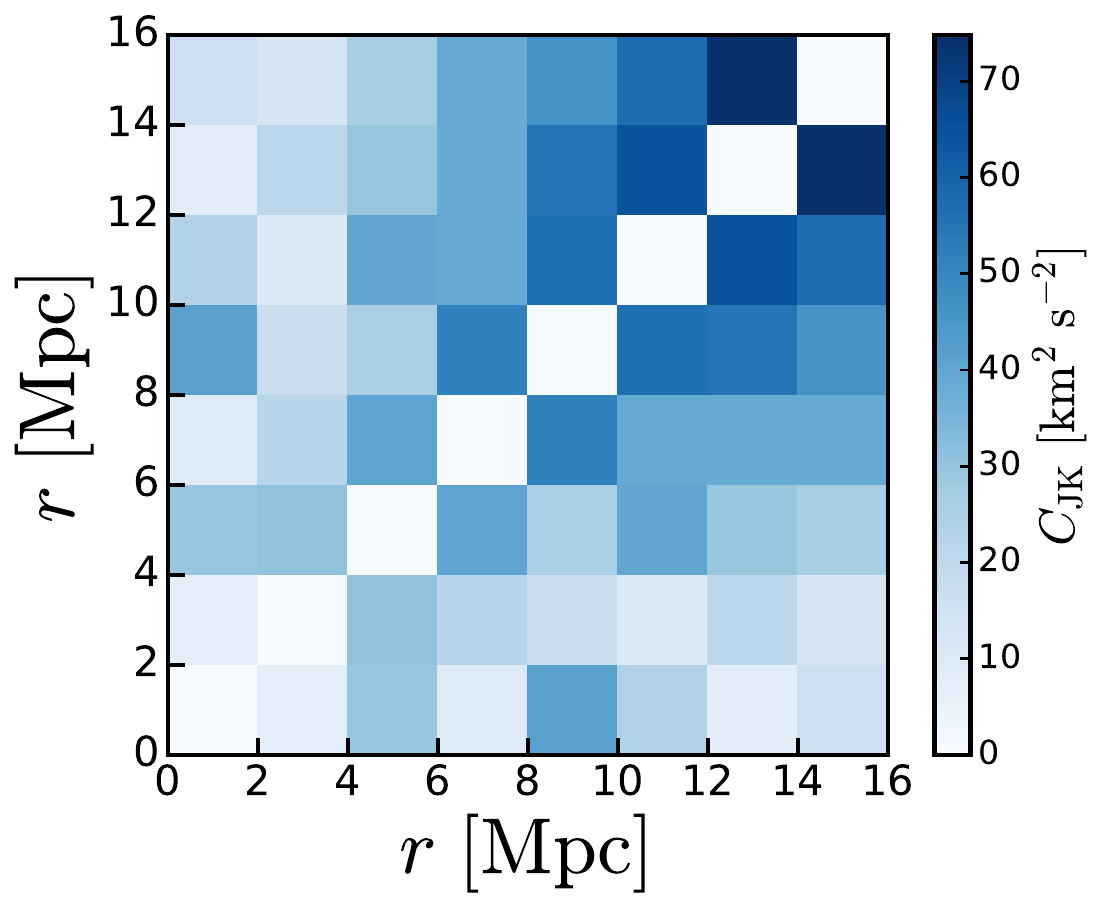}
  \caption{Error sources of $v_\mathrm{gg}$ and the same $r$ bins in different colored symbols are shifted horizontally slightly for clarity (left panel). The Covariance matrix of $C_\mathrm{JK}$ for different $r$ bins, with diagonal terms manually set to be 0 (right panel), where the colors show the magnitudes of $C_\mathrm{JK}$.}
  \label{fig:err_sources}
\end{figure*}

In this section, we present the error sources in galaxy-galaxy pairwise velocity $v_\mathrm{gg}$, as shown in Figure \ref{fig:err_sources}. All calculations in the following are based on $H_0=74.6\ \mathrm{km}\ \mathrm{s}^{-1}\ \mathrm{Mpc}^{-1}$ and $\Omega_m=0.27$ suggested by CF-4 \citep{tully2023}.

\subsubsection{Jackknife}

We use Jackknife by \texttt{Astropy} \citep{astropy:2013, astropy:2018, astropy:2022} to estimate the statistical covariance matrix of $v_\mathrm{gg}$, denoted as $C_\mathrm{JK}$,
\begin{equation}
    \left[C_\mathrm{JK}\right]_{ij} = \frac{N_\mathrm{gr}-1}{N_\mathrm{gr}} \sum_{k=1}^{N_\mathrm{gr}} \left[v_\mathrm{gg}^{(k)}(r_i) - \bar{v}_\mathrm{gg}^\mathrm{jack}(r_j)\right]^2,
\end{equation}
where $N_\mathrm{gr} = 10,014$ is the number of galaxies and $\bar{v}_\mathrm{gg}^\mathrm{jack}(r)\equiv\sum_{k=1}^{N_\mathrm{gr}}v_\mathrm{gg}^{(k)}(r)/N_\mathrm{gr}$. The term $v_\mathrm{gg}^{(k)}$ is derived from the $k^\mathrm{th}$ Jackknife sub-sample, incorporating all galaxies except the $k^\mathrm{th}$ one. The errors from the diagonal terms of $C_\mathrm{JK}$ are shown as blue squares in the left panel of Figure \ref{fig:err_sources}, while the off-diagonal terms are shown in the right panel.

\subsubsection{Peculiar velocity error}

When the line-of-sight peculiar velocity $v_p$ of a galaxy is calculated from Eq. (\ref{eq:peculiar-velocity}), we transfer the luminosity distance error $\Delta D_L$ on it (from the CF-4 grouped catalog \citep{tully2023}), denoted as $\Delta v_p$, and this is transferred to the final $v_\mathrm{gg}$, denoted as $\sigma_{vp}$. This error is reduced due to the relatively large number of pairs, especially in the large $r$ bins, shown as green circles in the left panel of Figure \ref{fig:err_sources}.

\subsubsection{Mass model}

In our pipeline, we first use Eqs. (\ref{eq:light-to-stellar-mass}) and (\ref{eq:stellar-to-halo-mass}) to determine the corresponding halo mass of a galaxy; then, we choose galaxies with halo masses within $[10^{11},10^{13}]\ M_\odot$ to calculate $v_\mathrm{gg}$.

Hence, here we consider the possible uncertainty caused by this procedure. First, $\log_{10}(M_g)$ in Eq. (\ref{eq:light-to-stellar-mass}) has around 0.1 dex scatter \citep{bell2003}, and also the coefficients in Eq. (\ref{eq:stellar-to-halo-mass}) have errors, i.e., $\Delta C_0=0.0006$, $\Delta(\log_{10}M_1)=0.025$, $\Delta\beta=0.0475$, and $\Delta\gamma=0.011$ \footnote{We take the mean of upper and lower one sigma, which is fair as the differences between them are small.} \citep{moster2010}. All of these can be propagated to the uncertainty of the final computed halo mass $\log_{10}M_h$, denoted as $\Delta(\log_{10}M_h)$.

Then we apply resampling 10,000 times. In each resampling, we generate a halo mass according to the Gaussian distribution centered at $\log_{10}M_h$ with width $\Delta(\log_{10}M_h)$, and we choose galaxies with halo masses within $[10^{11},10^{13}]\ M_\odot$ to recalculate $v_\mathrm{gg}$ accordingly.

Finally, we take the one sigma scatter of such resamplings, denoted as $\sigma_\mathrm{mass}$, to characterize the uncertainty of mass models we used, shown as orange diamond error bars in the left panel of Figure \ref{fig:err_sources}.

\subsubsection{Total error}

The total covariance matrix of $v_\mathrm{gg}$, denoted as $C_\mathrm{gg}$, combines contributions from all three sources discussed above, i.e., $[C_\mathrm{gg}]_{ij}=[C_\mathrm{JK}]_{ij} + (\sigma_{vp})^2\delta_{ij} + (\sigma_\mathrm{mass})^2\delta_{ij}$. The diagonal terms are shown as black hexagonal error bars in the left panel of Figure \ref{fig:err_sources}, which are mainly due to Jackknife.

In this section, although $H_0$ is set to 74.6 km s$^{-1}$ Mpc$^{-1}$ and $\Omega_m$ to 0.27, $\sigma_\mathrm{mass}$ is independent of $H_0$ and $\Omega_m$. In the following analysis, $\sigma_{vp}$ is recalculated for various $H_0$ and $\Omega_m$, assuming that $C_\mathrm{JK}$ does not change across these combinations.

\section{Results}\label{sc:fitting}

As discussed in Section \ref{sc:sub_ggpwv}, hereafter, the galaxy-galaxy pairwise velocity ($v_\mathrm{gg}$) and halo-halo pairwise velocity ($v_\mathrm{hh}$) are calculated over $r\in[0,16]$ Mpc using 8 linear bins.

\begin{figure*}
    \includegraphics[width=.95\textwidth]{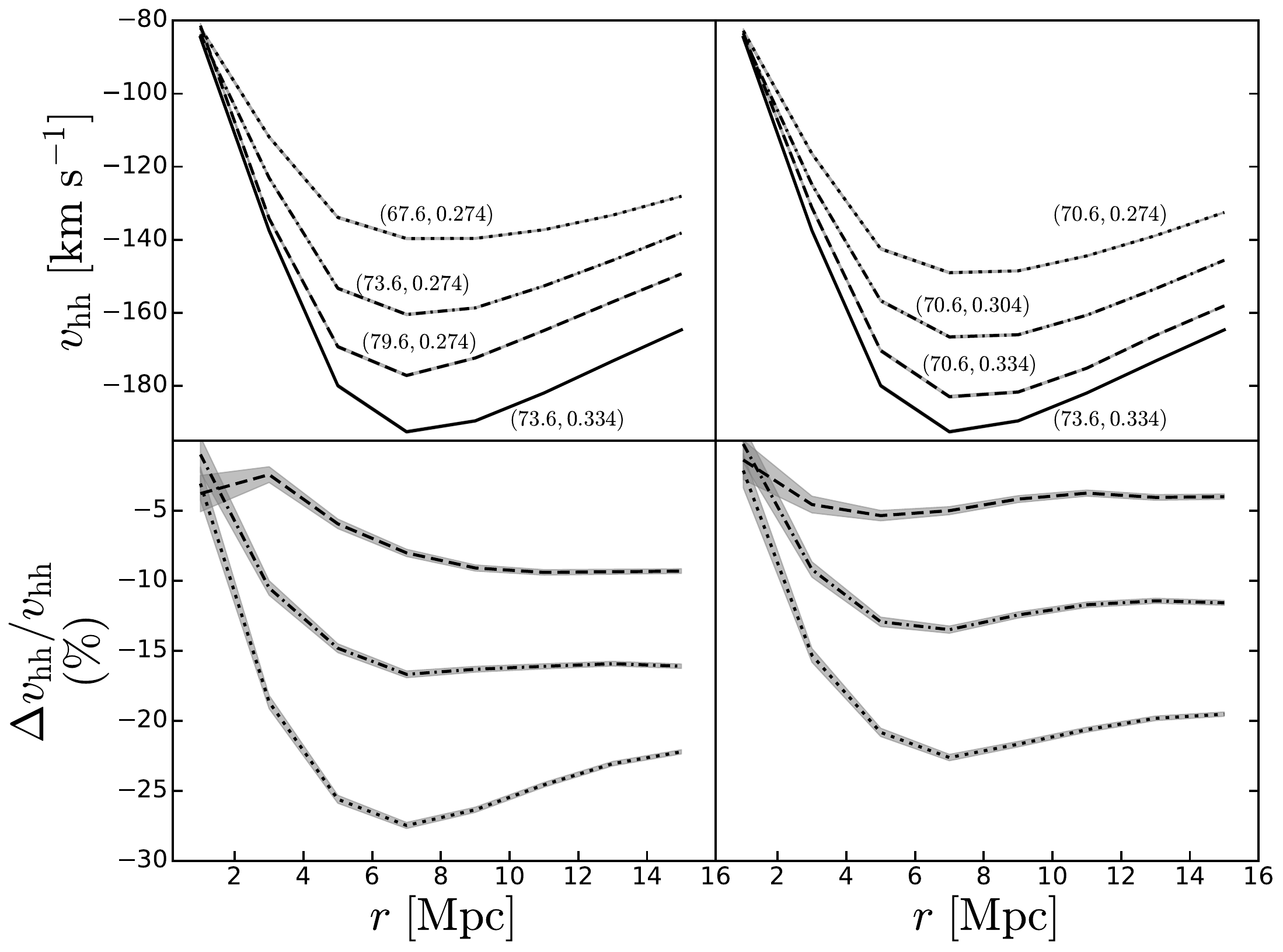}
    \caption{Halo-halo pairwise velocity $v_\mathrm{hh}$ for $\Omega_m=0.274$ and $H_0=$ 67.6 (dotted), 73.6 (dot-dashed), and 79.6 (dashed) (left panels), as well as $H_0=70.6$ and $\Omega_m=$ 0.274 (dotted), 0.304 (dot-dashed), and 0.334 (dashed) (right panels). The legends are for $(H_0,\Omega_m)$, where $H_0$ is in $\mathrm{km}\ \mathrm{s}^{-1}\ \mathrm{Mpc}^{-1}$. The fiducial case ($H_0=73.6$ and $\Omega_m=0.334$) is shown as solid lines in the top row. The bottom row shows the percentage deviations between the various styled lines and the fiducial case.}
    \label{fig:vhh_H0_Om}
\end{figure*}

The effects of $H_0$ and $\Omega_m$ on $v_\mathrm{hh}$ are shown in Figure \ref{fig:vhh_H0_Om}, which are degenerate, i.e., a larger value of either would lead to a larger magnitude of $v_\mathrm{hh}$. A larger $H_0$ with a fixed $\Omega_m$ means a higher matter density ($\Omega_mh^2$) and larger gravity, resulting in a larger magnitude of $v_\mathrm{hh}$. Similarly, increasing $\Omega_m$ while fixing $H_0$ would lead to not only a higher matter density but also a smaller dark energy density (assuming a flat universe), which in turn makes $v_\mathrm{hh}$ more negative. In particular, at $r\lesssim4$ Mpc, their impact on $v_\mathrm{hh}$ diminishes due to the reduced cosmological effects within the halo merger range.

We can apply the linear regression on the halo-halo pairwise velocity $v_\mathrm{hh}$ for each $r$ bin,
\begin{equation}\label{eq:fit_pwv_H0_Om}
\resizebox{1.03\columnwidth}{!} 
{$
    \begin{aligned}
	v_\mathrm{hh}(H_0,\Omega_m,r) &=  v_\mathrm{hh}(73.6,0.334,r) \\
        &\times\left[
        C_H(r) \frac{H_0 - 73.6}{73.6} + C_m(r) \frac{\Omega_m - 0.334}{0.334} + 1
        \right],
    \end{aligned}
$}
\end{equation}
where $C_H(r)$ and $C_m(r)$ are the $r$-dependent fitting parameters. Here, $H_0$ is in units of $\mathrm{km}\ \mathrm{s}^{-1}\ \mathrm{Mpc}^{-1}$. To evaluate the impacts of cosmic variance from cosmological simulations, we analyze $v_\mathrm{hh}$ across various $H_0$ and $\Omega_m$ using two different initial realizations. We find that although the cosmic variance causes notable changes in $v_\mathrm{hh}$ values for the same set of $H_0$ and $\Omega_m$ between two realizations, the effects of $H_0$ and $\Omega_m$ on $v_\mathrm{hh}$ remain consistent between different initial realizations within the relevant range. Therefore, one realization is adequate for extracting $C_H(r)$ and $C_m(r)$ in Eq. (\ref{eq:fit_pwv_H0_Om}). To reduce the impact of cosmic variance on the values of $v_\mathrm{hh}(73.6,0.334,r)$, we replace $v_\mathrm{hh}(73.6,0.334,r)$ by its mean of a total of 12 simulations mentioned in Section \ref{sc:simu_method}, $\overline{v}_\mathrm{hh}(73.6,0.334,r)$. More details are provided in Appendix \ref{sc:cosmic_variance}.

Then, a $\chi^2$ is defined as
\begin{equation}
    \chi^2(H_0,\Omega_m) = \sum_{ij} D_v(H_0,\Omega_m,r_i) C^{-1}_{ij} D_v(H_0,\Omega_m,r_j),
\end{equation}
where \small$D_v(H_0,\Omega_m,r_i)\equiv v_\mathrm{gg}(H_0,\Omega_m,r_i)-v_\mathrm{hh}(H_0,\Omega_m,r_i)$\normalsize. The covariance matrix $C_{ij} = [C_\mathrm{gg}]_{ij} + (\Delta\overline{v}_\mathrm{hh})^2\delta_{ij}$, where $\Delta\overline{v}_\mathrm{hh}$ is the uncertainty of the simulation model in the corresponding $r$ bin, which originates from the uncertainties of fitting parameters $C_m$ and $C_H$ in Eq. (\ref{eq:fit_pwv_H0_Om}). The calculations for the observational matrix $C_\mathrm{gg}$ are discussed in Section \ref{sc:error_analysis}.

We assume the likelihood function follows the Gaussian form
\begin{equation}\label{eq:likelihood}
    \mathcal{L}(H_0,\Omega_m) \sim e^{-\chi^2(H_0,\Omega_m)/2}p(H_0)p(\Omega_m),
\end{equation}
where $p(H_0)$ and $p(\Omega_m)$ are the uniform priors for $H_0$ and $\Omega_m$, with ranges [60, 80] $\mathrm{km}\ \mathrm{s}^{-1}\ \mathrm{Mpc}^{-1}$ and [0.2, 0.4], respectively. 

We use the Markov Chain Monte Carlo (MCMC) method to apply the fitting, which is employed by \texttt{emcee} \citep{emcee2013}. At each step, given $(H_0,\Omega_m)$, $v_\mathrm{hh}$ is calculated using Eq. (\ref{eq:fit_pwv_H0_Om}), while $v_p$ and positions of galaxies are recalculated by Eqs. (\ref{eq:peculiar-velocity}) and (\ref{eq:gr_pos}), respectively. Then, $v_\mathrm{gg}$ is calculated by Eq. (\ref{eq:galaxy_pwv}). With the knowledge of $v_\mathrm{hh}$ and $v_\mathrm{gg}$, we calculate the likelihood $\mathcal{L}$ by Eq. (\ref{eq:likelihood}).

\begin{figure*}
\includegraphics[width=.45\textwidth]{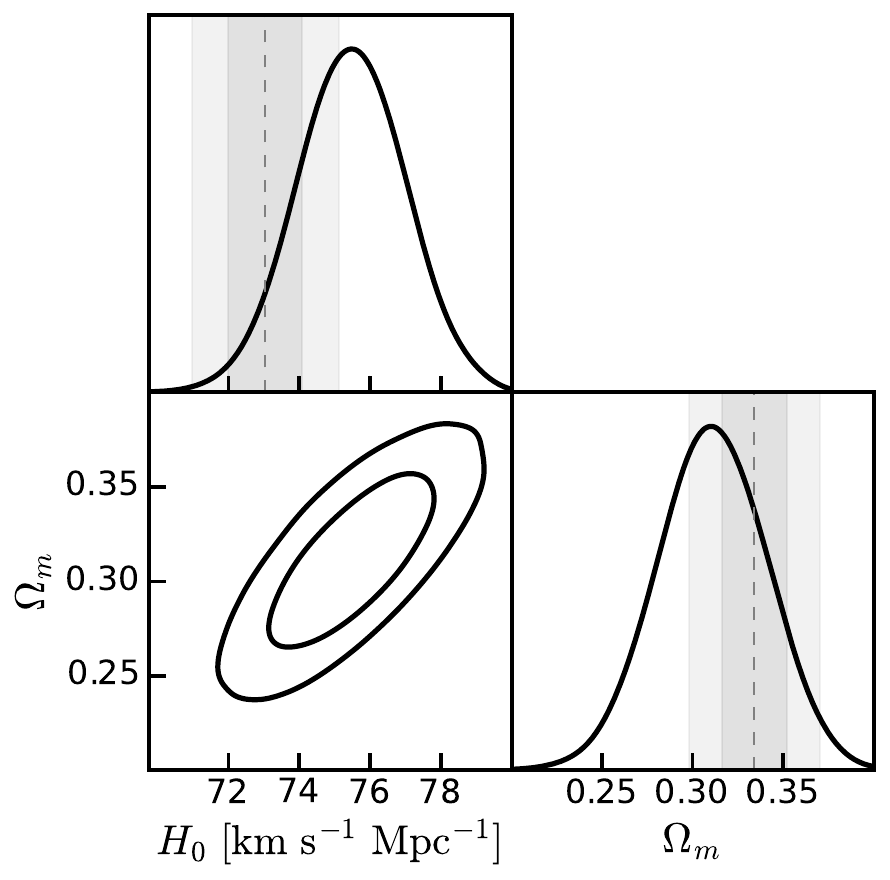}
\includegraphics[width=.45\textwidth]{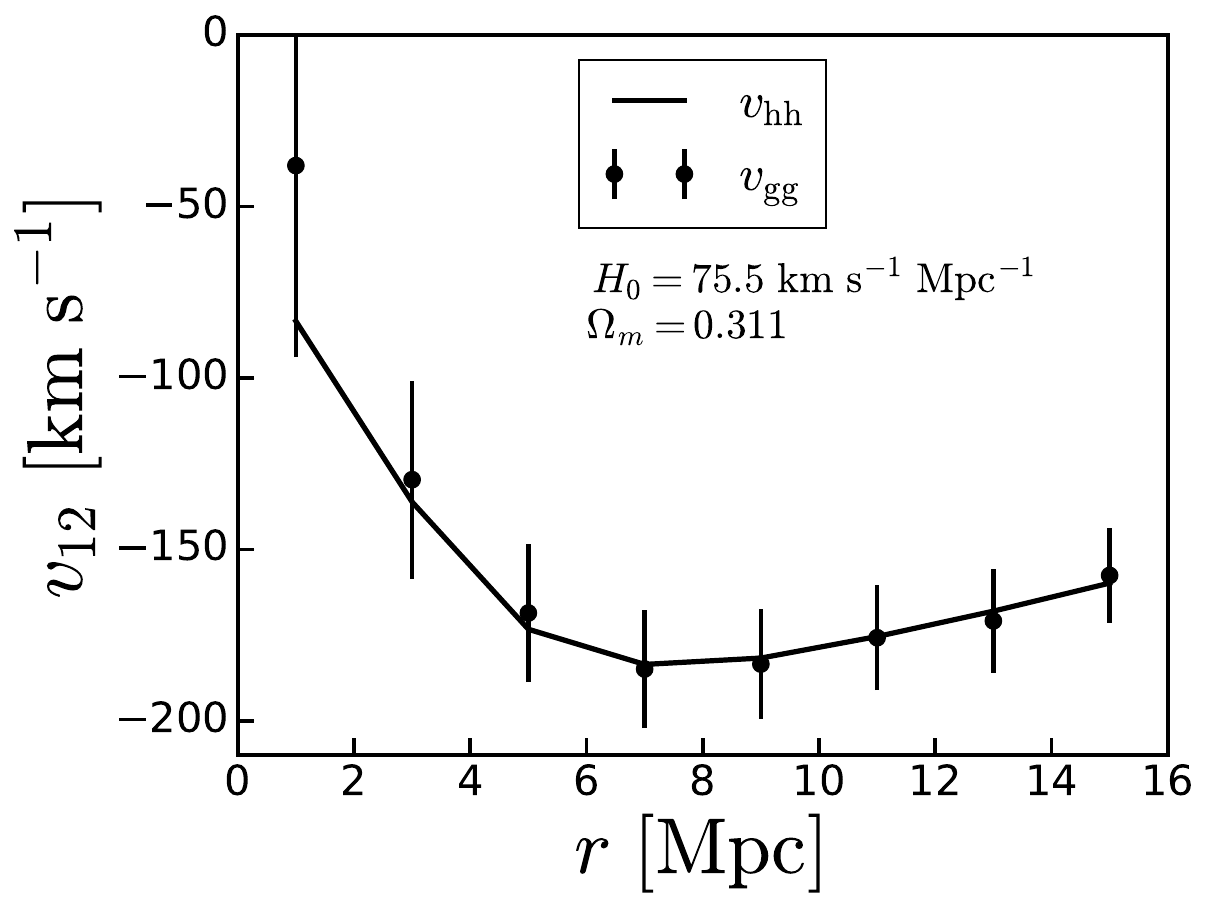}
\caption{Contours (at 68\% and 95\% CL) and PDFs for $H_0$ and $\Omega_m$ (left panel). The gray bands in $H_0$ and $\Omega_m$ PDFs are from SH0ES ($H_0=73.04\pm1.04\ \mathrm{km}\ \mathrm{s}^{-1}\ \mathrm{Mpc}^{-1}$, \citealt{SHOES2022}) and Pantheon+ ($\Omega_m=0.334\pm0.018$, \citealt{Pantheon2022}), respectively. Halo-halo pairwise velocity $v_\mathrm{hh}$ (solid line, from simulation) and galaxy-galaxy pairwise velocity $v_\mathrm{gg}$ (data points with error bars, from observation) when taking $H_0$ and $\Omega_m$ at the mean values of the MCMC results are shown in the right panel.}
\label{fig:fit_H0_Om_best_fit}
\end{figure*}

Our MCMC results are shown in the left panel of Figure \ref{fig:fit_H0_Om_best_fit}, and the mean values with 68\% confidence level (CL) are $H_0=75.5\pm1.4\ \mathrm{km}\ \mathrm{s}^{-1}\ \mathrm{Mpc}^{-1}$ and $\Omega_m=0.311^{+0.029}_{-0.028}$, which are consistent with the SH0ES result ($H_0=73.04\pm1.04\ \mathrm{km}\ \mathrm{s}^{-1}\ \mathrm{Mpc}^{-1}$, \citealt{SHOES2022}) and both Planck ($\Omega_m=0.3153\pm0.0073$, \citealt{aghanim2020planck}) and Pantheon+ results ($\Omega_m=0.334\pm0.018$, \citealt{Pantheon2022}), respectively. The current uncertainty on $H_0$ ($\Omega_m$) is 2\% (9\%). The right panel of Figure \ref{fig:fit_H0_Om_best_fit} shows $v_\mathrm{hh}$ (from simulation) and $v_\mathrm{gg}$ (from observation) at the mean values of $H_0$ and $\Omega_m$.

Moreover, we can also estimate the precision we can achieve if the Jackknife and peculiar velocity errors are negligible, as discussed in Section \ref{sc:error_analysis}. Then, the projected uncertainties would be $(+0.43,-0.42)\ \mathrm{km}\ \mathrm{s}^{-1}\ \mathrm{Mpc}^{-1}$ for $H_0$ and $(+0.0073,-0.0074)$ for $\Omega_m$, or 0.6\% and 2\%, respectively.

\section{Conclusions}\label{sc:summary}

In this Letter, we measure $H_0$ and $\Omega_m$ using the mean galaxy pairwise peculiar velocity \footnote{Here, ``mean" is the arithmetic average of all pairs of galaxies with the same separations $r$.} in the nonlinear and quasi-linear region for the first time. By applying MCMC to fit simulation models with the galaxy data drawn from the CF-4 grouped catalog, we find that 

\hspace*{\fill}

(i) we can achieve 2\% and 9\% measurements in $H_0$ and $\Omega_m$, respectively, and obtain $H_0=75.5\pm1.4\ \mathrm{km}\ \mathrm{s}^{-1}\ \mathrm{Mpc}^{-1}$ and $\Omega_m=0.311^{+0.029}_{-0.028}$ with 68\% CL. The value of $H_0$ we obtain is consistent with the SH0ES result \citep{SHOES2022}, and our $\Omega_m$ agrees with both the Planck \citep{aghanim2020planck} and Pantheon+ results \citep{Pantheon2022};

(ii) if, in the future, the statistical errors become negligible, we can get much more precise measurements of $H_0$ and $\Omega_m$, with 0.6\% and 2\% uncertainty, respectively.

\begin{acknowledgements}
We thank Renbin Yan, Yin-Zhe Ma, and R. Brent Tully for discussions, and the anonymous referee for helpful comments. The computational resources used for the simulations in this work were kindly provided by the Chinese University of Hong Kong Central Research Computing Cluster. This research is supported by grants from the Research Grants Council of the Hong Kong Special Administrative Region, China, under Project No.s AoE/P-404/18 and 14300223. S.L. acknowledges the support by the National Natural Science Foundation of China (NSFC) grant (no. 12473015). H.H. is supported by the China Postdoctoral Science Foundation grant No. 2024M763213. All plots in this Letter are generated by \texttt{Matplotlib} \citep{matplotlib2007} and \texttt{Getdist} \citep{getdist2019}, together with \texttt{SciPy} \citep{scipy2020}, \texttt{NumPy} \citep{numpy2020}.
\end{acknowledgements}

\appendix

\section{Tests on the choice of \texorpdfstring{$\lowercase{v_{p}^{\rm max}}$}{vpmax}}\label{sc:vp_cut}

We estimate the line-of-sight peculiar velocity $v_p$ of a galaxy using the halo catalog from the \texttt{Uchuu} simulation \citep{ishiyama2021Uchuu}, selecting host halos within the same virial mass range $[10^{11},10^{13}]\ M_\odot$ as described in the main text, obtaining a total of 223,900,197 host halos. We randomly generate three observers inside the \texttt{Uchuu} simulation box and calculate the corresponding $v_p$ of the host halos. The $v_p$ histograms follow Gaussian distributions with $3\sigma=960.5,963.2,960.3\ \mathrm{km}\ \mathrm{s}^{-1}$, respectively. Based on this, taking $H_0=74.6\ \mathrm{km}\ \mathrm{s}^{-1}\ \mathrm{Mpc}^{-1}$ and $\Omega_m=0.27$ \footnote{Although $\Omega_m$ differs from our final MCMC results, we have confirmed that our results remain practically unchanged for the values of $\Omega_m$ between 0.27 and 0.32.} as suggested by CF-4 \citep{tully2023}, we apply a bound on the peculiar velocity calculated from Eq. (\ref{eq:peculiar-velocity}), $|v_p|\leq v_p^\mathrm{max}$, with $v_p^\mathrm{max}=1000\ \mathrm{km}\ \mathrm{s}^{-1}$. This leaves us with 10,014 galaxies, which we analyze in the main text.

\begin{figure*}
\centering
    \includegraphics[width=.75\textwidth]{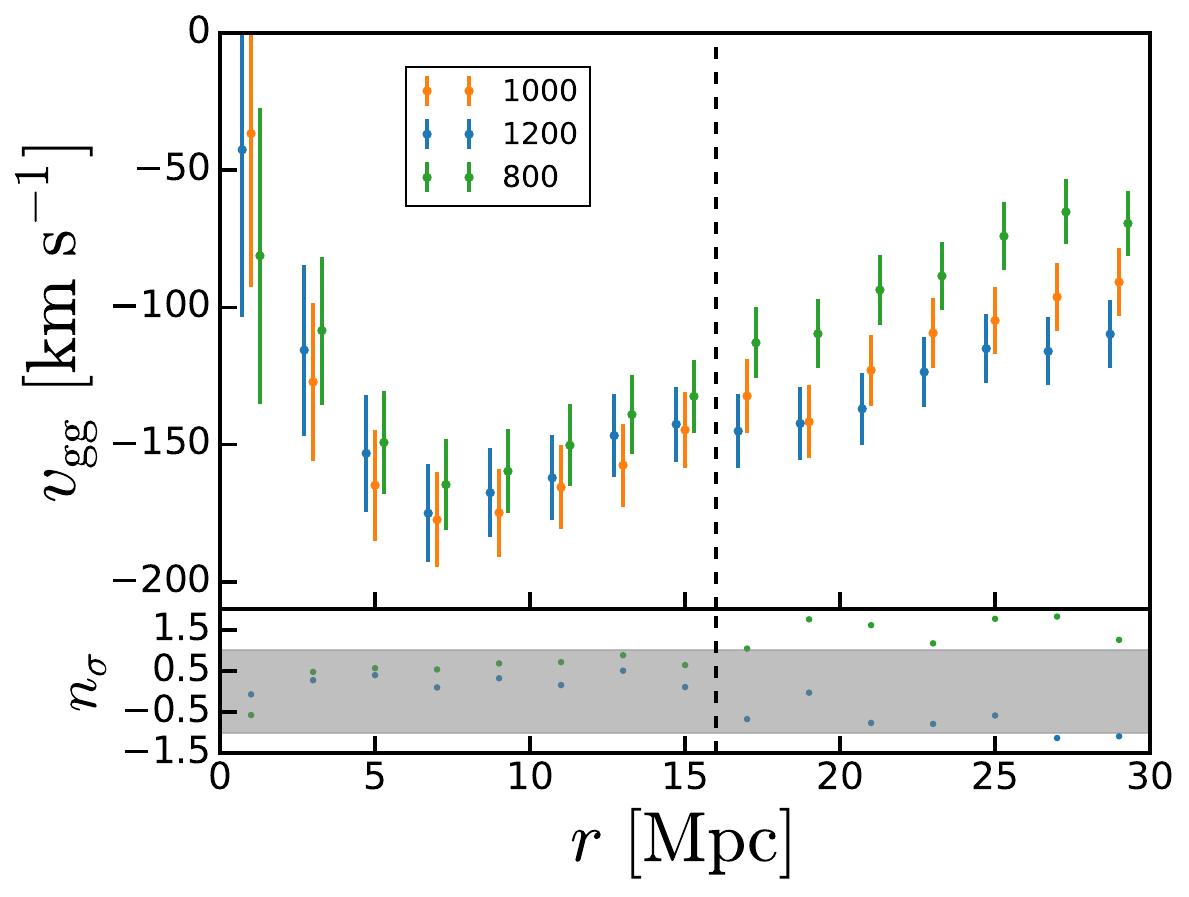}
    \caption{Galaxy-galaxy pairwise velocity for various $v_p^\mathrm{max}$ (in $\mathrm{km}\ \mathrm{s}^{-1}$) represented by different colors, with both statistical and systematic error bars (upper subpanel). For clarity, the identical $r$ bins depicted in different colors are shifted horizontally slightly. The lower subpanel shows the deviations of $v_p^\mathrm{max}=$ 800 and 1200 km s$^{-1}$ cases from $v_p^\mathrm{max}=1000$ km s$^{-1}$, quantified in sigma units ($n_\sigma$). The gray band denotes $\pm1\sigma$ deviation. The black dashed line indicates $r=16$ Mpc.}
    \label{fig:vp_cut}
\end{figure*}

Then, we examine how the choice of $v_p^\mathrm{max}$ affects $v_\mathrm{gg}$ by adjusting $v_p^\mathrm{max}$ around 1000 km s$^{-1}$. We sample 15 linear bins within $r\in[0,30]$ Mpc, calculating $v_\mathrm{gg}$ for $v_p^\mathrm{max}=$ 800, 1000, and 1200 km s$^{-1}$. The results are shown in the upper subpanel of Figure \ref{fig:vp_cut}, while the lower subpanel presents the deviation of $v_\mathrm{gg}$ for different $v_p^\mathrm{max}$ values from that for $v_p^\mathrm{max}=1000$ km s$^{-1}$, quantified as the number of sigmas, assuming Gaussian distributions,
\begin{equation}
    n_\sigma(r) \equiv 
    \frac
    {  \left.v_\mathrm{gg}(r)\right\vert_{v_{p,1}^\mathrm{max}} 
     - \left.v_\mathrm{gg}(r)\right\vert_{v_{p,0}^\mathrm{max}}
    }
    {\sqrt{
            \left.\Delta v^2_\mathrm{gg}(r)\right\vert_{v_{p,1}^\mathrm{max}} 
          + \left.\Delta v^2_\mathrm{gg}(r)\right\vert_{v_{p,0}^\mathrm{max}} 
          } 
    },
\end{equation}
where $v_{p,1}^{\mathrm{max}}=$ 800 or 1200 km s$^{-1}$, while $v_{p,0}^{\mathrm{max}}=$ 1000 km s$^{-1}$. The error of $v_\mathrm{gg}$, denoted as $\Delta v_\mathrm{gg}$, is evaluated as per Section \ref{sc:error_analysis}. We find that $v_\mathrm{gg}$ is not sensitive to the exact value of $v_p^\mathrm{max}$ for $r\lesssim16\ \mathrm{Mpc}$, i.e., $|n_\sigma|\lesssim1$.

Consequently, throughout this Letter, we calculate $v_\mathrm{hh}$ and $v_\mathrm{gg}$ using 8 linear bins within $r\in[0,16]$ Mpc. We also check the effects of binning numbers, by repeating the analysis in the main text using 4 and 12 linear bins within this range and find that the constraints on $H_0$ and $\Omega_m$ are not affected.

\section{Cosmic variance}\label{sc:cosmic_variance}

\begin{figure*}
  \centering
  \includegraphics[width=.75\textwidth]{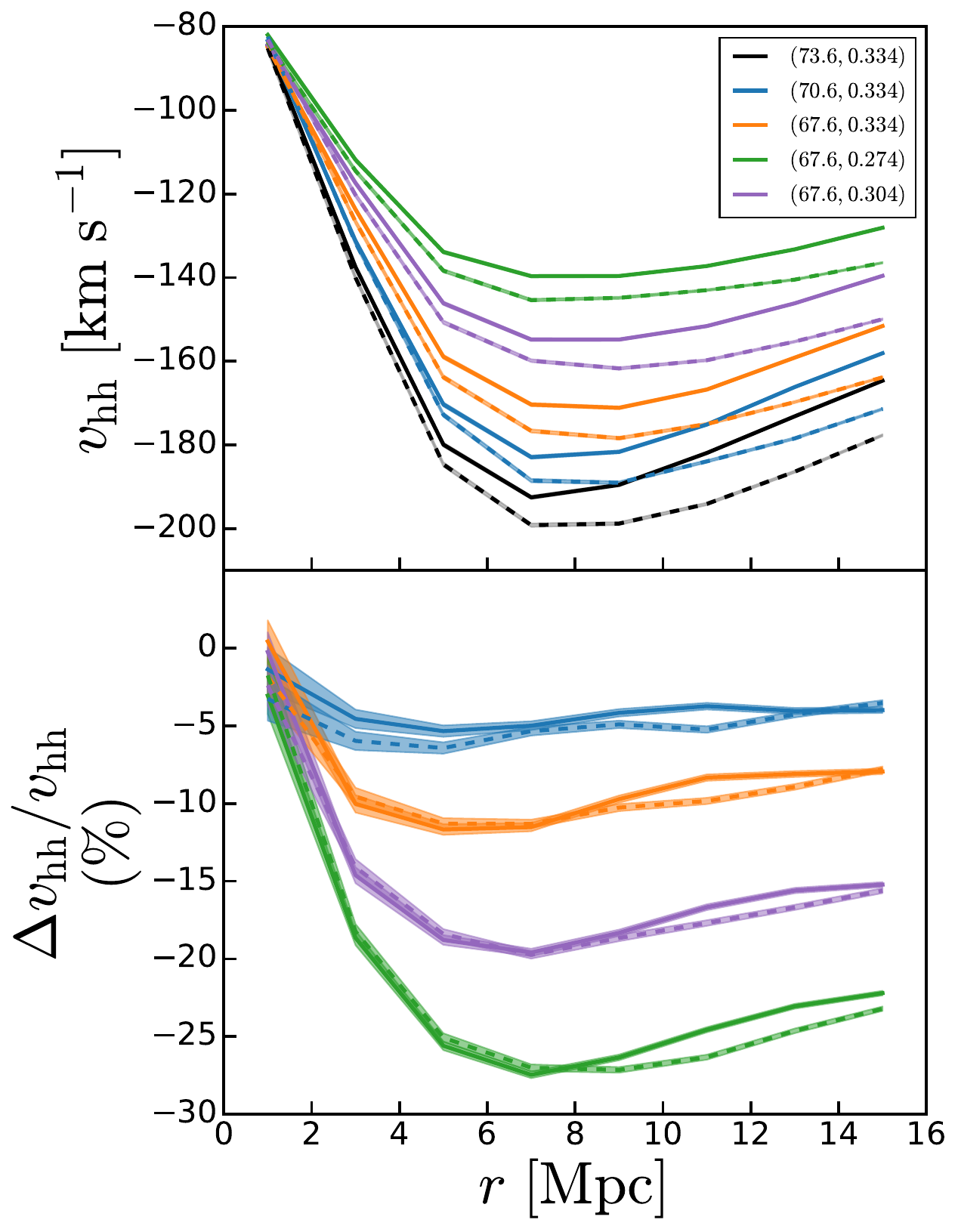}
  \caption{Halo-halo pairwise velocity ($v_\mathrm{hh}$, upper panel) and percentage deviations of halo-halo pairwise velocity with respect to the fiducial case ($H_0=73.6$ and $\Omega_m=0.334$) (lower panel), for two different initial random seeds (solid and dashed lines). The shaded bands indicate the 1$\sigma$ statistical error. The legends are for ($H_0,\Omega_m$), where $H_0$ is in $\mathrm{km}\ \mathrm{s}^{-1}\ \mathrm{Mpc}^{-1}$.}
  \label{fig:cosmic_variance}
\end{figure*}

In Figure \ref{fig:cosmic_variance}, we show $v_\mathrm{hh}$ for various ($H_0$, $\Omega_m$) combinations derived from two distinct initial realizations (solid and dashed lines). While the values of $v_\mathrm{hh}$ exhibit significant variations between realizations due to the cosmic variance (upper panel of Figure \ref{fig:cosmic_variance}), the effects of $H_0$ and $\Omega_m$, expressed as changes of halo-halo pairwise velocity $v_\mathrm{hh}$ relative to the fiducial case (Eq. (\ref{eq:fit_pwv_H0_Om})), remain consistent across different initial seeds (lower panel of Figure \ref{fig:cosmic_variance}).

Consequently, in the main text, we only use one random seed to estimate the effects of $H_0$ and $\Omega_m$, and we average 12 simulations for the fiducial case ($H_0=73.6\ \mathrm{km}\ \mathrm{s}^{-1}\ \mathrm{Mpc}^{-1}$ and $\Omega_m=0.334$) to reduce the effects of cosmic variance.

\end{document}